\documentclass{mem}

\usepackage{txfonts}
\usepackage{graphicx}
\usepackage{graphics}
\usepackage{epsf,epsfig}
\usepackage[dvips]{color}

\usepackage{rotate}
\usepackage{color}
\usepackage{amssymb,amsfonts}
\usepackage{latexsym}

\usepackage[a4paper]{hyperref}

\idline{75}{282}

\begin{document}

\def\teff{$T\rm_{eff }$}
\def\kms{$\mathrm {km s}^{-1}$}

\title{
Hard X-rays from Galaxy Clusters \& SIMBOL-X}

\subtitle{}

\author{
G. Brunetti\inst{1}, R. Cassano\inst{1,2}, G. Setti\inst{1,2}
}

\offprints{G. Brunetti}

\institute{
Istituto Nazionale di Astrofisica --
Istituto di Radioastronomia, Via P. Gobetti 101,
I-40129 Bologna, Italy
\and
Universita' di Bologna, Dipartimento di Astronomia,
Via Ranzani 1, I-40127 Bologna, Italy
\email{brunetti@ira.inaf.it}
}

\authorrunning{Brunetti et al.}

\titlerunning{HXR from Galaxy Clusters}

\abstract{
Non thermal emission from galaxy clusters demonstrates the existence
of relativistic particles and magnetic fields in the Intra Cluster
Medium (ICM). Present instruments do not allow to firmly establish the
energy associated to these components.
In a few years gamma ray observations will put important constraints
on the energy content of non thermal hadrons in clusters, 
while the combination of radio and hard X-ray data will be
crucial to measure the energy content in the form of relativistic
electrons and magnetic field.
SIMBOL-X is expected to drive an important breakthrough in the field
also because it is expected to operate in combination with the
forthcoming low frequency radio telescopes (LOFAR, LWA).
In this contribution we report first estimates of {\it statistical properties}
of the hard X--ray emission in the framework of the 
{\it re-acceleration model}.
This model allows to reproduce present radio data for Radio Halos
and to derive expectations for future low frequency radio observations,
and thus our calculations provide hints for observational strategies
for future radio and hard--X-ray combined observations.

\keywords{
acceleration of particles - radiation mechanisms: non--thermal -
galaxies: clusters: general - X--rays: general
}}

\maketitle{}

\section{Introduction}

Clusters of galaxies represent the largest virialized 
structures in the present Universe.
Rich clusters have typical total masses of 
$10^{15} M_{\odot}$, mostly in the form of dark matter, while
$\sim 5\%$ of the mass 
is in the form of a hot ($T \sim 10^8 K$), tenuous ($n_{gas} \sim
10^{-3}-10^{-4} cm^{-3}$), X-ray emitting gas. 
In terms of energy density, the gas is typically heated to
roughly the virial temperature, but there is also room to accomodate
a non-negligible amount of non-thermal energy.

Clusters are ideal astrophysical environments
for particle acceleration and cosmic rays
(CR) accelerated within the cluster volume
are expected to be confined for cosmological times
(e.g., Blasi, Gabici, Brunetti 2007, BGB07, for a review).
The bulk of the energy of these CRs is expected in
protons since they have radiative and collisional life--times
much longer than those of the electrons.
While present gamma ray observations can only provide
upper limits to the average energy density of
CR protons in the ICM (e.g. Reimer et al. 2004),
evidence of a non-thermal component is in fact obtained from
radio observations of a fraction of galaxy clusters
showing synchrotron emission 
on Mpc scales : Radio Halos, fairly symmetric sources at the cluster
center, and Radio Relics, elongated sources at the cluster periphery 
(e.g., Feretti 2005).

Although the bulk of present data comes from radio observations, 
theoretically a substantial fraction of the 
non thermal radiation is expected from inverse Compton (IC)
scattering of the photons of the cosmic
microwave background (e.g., Sarazin 1999).
Measuring IC emission from clusters in the hard X--rays
is extremely important to derive the energy density
of emitting electrons and 
the strength of the magnetic field when these measures
are combined with radio data.
Despite the poor sensitivity of present and past hard X--ray telescopes,
several groups have claimed detection of hard X--ray emission (HXR) 
in a few massive clusters (e.g., Fusco-Femiano et al.~2004;
Petrosian et al.~2006; Rephaeli et al.~2006; see also Rossetti \&
Molendi 2004 and Fusco-Femiano et al.~2007 for a discussion
on the strength of the HXR detection in the Coma cluster).

Thanks to its sensitivity and capability to perform hard X--ray imaging
SIMBOL--X will open a new era in the study of non thermal
radiation from galaxy clusters.
In this contribution we report first expectations on the 
Luminosity Functions (LFs) and
number counts of HXR from clusters.
We calculate only the contribution to the IC spectrum from electrons 
re-accelerated by turbulence in the ICM which are 
the responsible for the origin of Radio Halos in the context of
the {\it re-acceleration scenario}.

\section{HXR expectations from the re-acceleration scenario}

\subsection{Introduction}

Mpc scale radio emission at the level of 
Radio Halos is found in a fraction of massive and
merging galaxy clusters (e.g., Feretti 2005).
The connection between 
Radio Halos and cluster mergers, the very large extension of 
Radio Halos and their complex spectral properties
pose serious challenges to our understanding of these
sources, at least when a quantitative comparison between models 
and data is performed.

\noindent
A promising possibility to explain Radio Halos is given by the
{\it re--acceleration scenario} (Brunetti et al. 2001; Petrosian 2001).
In this scenario particles are supposed to be re--accelerated on
large scales by
MHD turbulence injected in the ICM during cluster--cluster mergers.
Although the details of the
physics of turbulence and of
stochastic particle acceleration are still poorly understood,
detailed calculations of Alfv\'enic and magnetosonic acceleration 
suggest taht efficient turbulent acceleration may take place
in the ICM (Brunetti et al. 2004; Brunetti \& Lazarian 2007).

\subsection{A statistical approach}

In Cassano \& Brunetti (2005) and Cassano et al.(2006) we have 
modeled the statistical properties of Radio Halos as expected
in the {\it re-acceleration scenario}.
In these papers we derive the merging history of a large synthetic 
population of galaxy clusters, the turbulence injected during mergers 
and follow the process of stochastic acceleration of the
relativistic electrons driven by this turbulence.

This allows us to get semi-analytic expectations
for the LFs of Radio Halos assuming a $Mpc^3$ emitting region.
These are given as :

\begin{equation}
{dN_{H}(z,\nu)\over{dV\,dP_{\nu}}}=
n_{PS}\times
{\cal P}_{\Delta z}^{\Delta M}(\nu) \left( {dP_{\nu}\over dM} \right)^{-1}
\label{RHLF}
\end{equation}

where $n_{PS}=n_{PS}(M,z)$ is the {\it Press \& Schechter} mass
function, ${\cal P}_{\Delta z}^{\Delta M}(\nu)$ is the probability
to have Radio Halos emitting at frequency $\nu$
as measured in the population of
synthetic clusters 
and $dP_{\nu}\over dM$ can be derived from 
the radio power cluster -- mass correlation.
The statistical behaviour of ${\cal P}_{\Delta z}^{\Delta M}(\nu)$ 
depends on the magnetic field
strength in the emitting region and on its scaling with
cluster mass. The important point here is that
it has been shown that present radio data
can be reproduced provided that the magnetic field in the
emitting region is within some
allowed region in the $B$--$b$ plane, where $B(M) \propto M^b$
and $M$ is the virial mass of clusters (CBS06).

\subsection{IC Hard X-ray emission}

Starting from 
our previous statistical calculations for Radio Halos
we can obtain simple estimates of the
LFs of HXR as:

\begin{equation}
{dN_{HXR}(z,P_{hxr})\over{dV\,dP_{hxr}}}=
{dN_{H}(z,P_{150})\over{dV\,dP_{150}}}
\left( {{dP_{150} }\over{
dP_{hxr} }} \right)
\label{HXRLF}
\end{equation}

where the ratio between IC and synchrotron power depends on $B$:

\begin{equation}
\nu_{hxr} d P_{hxr} \approx 
\left( {{3.2 (1+z)^2 }\over{B_{\mu G}}} \right)^2 
d P_{150} \nu_{150}
\label{PP}
\end{equation}

\noindent
and where the HXRs are anchored to the synchrotron emission
at $\nu=$150 MHz 
as the electrons emitting around this frequency in $\mu$G fields are
also the responsible for the IC emission in the hard X--rays.

Eqs.\ref{RHLF}--\ref{PP} 
allow to obtain non K--corrected LFs for HXR.
It should also be mentioned that these estimates provide 
lower limits (within a factor of $\approx 3$)
to the HXRs since a substantial fraction of these HXRs 
is expected to come from regions external to the central
Mpc$^3$ (Brunetti et al. 2001,04; Colafrancesco et al. 2005).

The magnetic field is a crucial parameter in our calculations.
Rotation Measures (RM) give values of the 
field of several
$\mu G$, but these estimates are affected by uncertainties
in the topology of the
field and the spatial distribution of the thermal electrons, as well as by
the subtraction of the intrinsic RM at the source (e.g., Govoni \&
Feretti 2004 for a review).
Smaller fields, of the order of a few tenths
of $\mu G$, are obtained by detection of HXR (e.g., Fusco-Femiano
et al.~2004).
This latter method relies on the assumption that the
diffuse radio emission and the HXRs are cospatial and
produced by the same population of relativistic electrons via
synchrotron and IC respectively .

Although many theoretical attempts have shown that this discrepancy may be
alleviated by considering the radial profile of the magnetic field strength
in clusters, the correct shape of the spectrum of the emitting electrons,
and possible anisotropic effects (e.g., Brunetti et al.~2001; 
Petrosian 2001), deep hard X-ray
observations should be able to definitely solve this point.

\begin{figure}
\resizebox{\hsize}{!}{\includegraphics[clip=true]{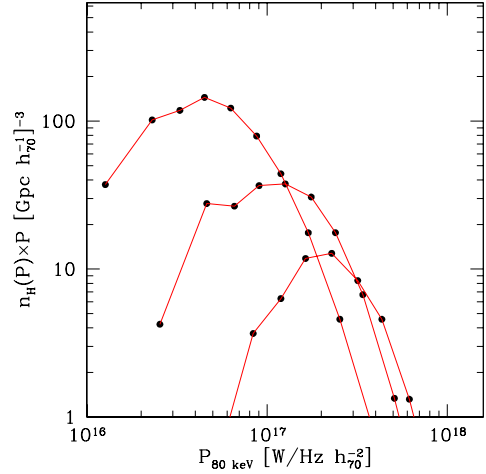}}
\caption{\footnotesize
LFs of HXRs in clusters. From top to bottom
one has z=0.05, 0.25, 0.45 .
Calculations are obtained assuming $B(M) \propto M^b$
with $B(<M>)=0.2 \mu$G, $<M>=1.6\times 10^{15}$M$_{\odot}$,
and $b=0.6$; this configuration is within the
allowed region to match the radio--X-ray correlations of 
Radio Halos with the {\it re-acceleration model} (CBS06).
}
\label{eta}
\end{figure}
\begin{figure}[]
\resizebox{\hsize}{!}{\includegraphics[clip=true]{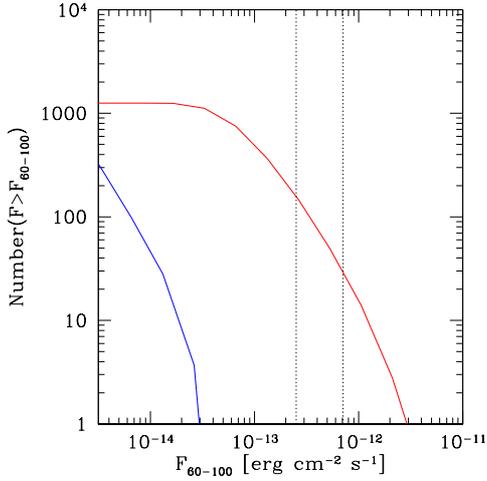}}
\caption{
\footnotesize
Number counts of HXR.
Right curve is obtained for the parameters in Fig.~1, Left
curve is obtained for $b=1.5$ and $B(<M>)=2 \mu$G.
Dotted lines give a reference range for SIMBOL-X sensitivity.
}
\label{li_vhel}
\end{figure}

\begin{figure}[]
\resizebox{\hsize}{!}{\includegraphics[clip=true]{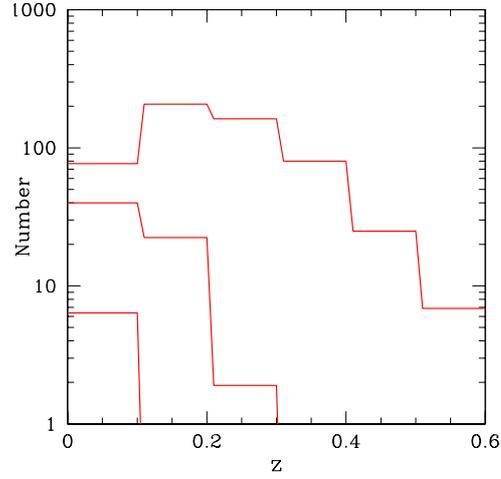}}
\caption{
\footnotesize
Number of clusters with detectable HXR as a function of redshift.
From top to bottom we assumed the following sensitivities :
F$_{60-100}$[erg cm$^{-2}$ s$^{-1}$]=
10$^{-13}$, 5$\times$ 10$^{-13}$, 10$^{-12}$.
Calculations are obtained by assuming the parameters in Fig.~1
}
\label{li_vhel}
\end{figure}

In Fig.~1 we report expected 
LFs of HXRs at different redshifts assuming a 
low value of the magnetic field in the ICM (see caption),
in which case the IC luminosities are maximized.
The flattening/cut--off at smaller
luminosities is due to the fact that at these luminosities the LFs
are contributed by clusters with masses $\leq 10^{15}$M$_{\odot}$
in which case the particle 
re-acceleration is less efficient (CBS06).

In Fig.~2 we report expected number counts by assuming two 
scenarios for the magnetic field in the emitting region
(see caption).
The important point here is that SIMBOL-X is expected to discover
HXRs in $\approx 30-100$ clusters in the case of low $B$, while 
only upper limits to the presence of HXRs will be obtained 
for large values of $B$.
This demonstrates the importance of future SIMBOL-X observations.

Finally, in Fig.~3 we report the redshift distribution of HXRs in the
Universe for different sensitivity levels and assuming the case of 
low magnetic field.
In this case SIMBOL-X is expected to discover HXRs in a few 
clusters with redshift $\approx 0.2$, while the bulk of detectable
HXRs is expected at lower redshift.

\section{Conclusions}

In this contribution we report first expectations for HXRs from
galaxy clusters, a more detailed study will be reported in a forthcoming
paper.
Calculations are performed in the framework of the 
{\it re-acceleration model} assuming
physical parameters which allow the {\it re-acceleration model} to
reproduce present data of the statistical behaviour 
of giant Radio Halos.

The strength of the magnetic field in the ICM is a crucial parameter
in our calculations and we have shown that SIMBOL-X will provide 
unique constraints. By assuming a value of the magnetic field averaged
in Mpc$^3$ volume of $\approx$0.2$\mu$G 
we find that SIMBOL-X will discover
HXRs in $\approx$30--100 clusters at z$\leq$0.2 .

\begin{acknowledgements}
We acknowledge partial support through grants ASI-INAF I/088/06/0 and
PRIN-MUR 2006-02-5203.
\end{acknowledgements}

\bibliographystyle{aa}

\end{document}